\documentclass[preprint,preprintnumbers]{revtex4}
\usepackage{graphicx,psfrag}

\begin{document}

\title{A MOND PROGRAMME FROM EINSTEIN HILBERT ACTION}
\author{Saneesh Sebastian}
\email{sebastiansaneesh@gmail.com}
\author{V C Kuriakose}
\email{vck@cusat.ac.in}

\begin{abstract}
In the usual derivation of Einstein's equation from action, the
surface terms are neglected. Hawking \cite{hw} gave a derivation of
the gravitational Hamiltonian keeping all surface terms. Using such
surface terms Easson et.al.\cite{ea} showed that Friedmann equation
could get modified and using the modified Friedmann equation they
could explain the cosmic acceleration. We study the effect of
surface terms on a galactic scale and find that  the classical limit
of the modified Friedmann equation will lead to a MOND like
acceleration term.

\end{abstract}

\maketitle

\address{Department of Physics, Cochin University of Science and
Technology, Cochin 682022, India.}

        Two of the challenging problems in galactic
 dynamics and modern cosmology are the flat rotation curve of galaxies and the acceleration of the universe.
  According to Newton's law of gravitation, the centrifugal acceleration $v^{2}/r$ should balance the
  gravitational attraction $GM(r)/r^{2}$ which immediately gives
  $v^{2}=GM(r)/r$. The velocities of stars as well as those of galaxies in galactic
clusters are not
 following the predicted Newtonian $r^{-1}$ law.The velocity approximately levels off
 with r in the halo region. The only way to interpret this result of observation
  is to accept that the mass M(r) increases linearly with distance r. Luminous
  mass distribution in the galaxy does not follow this behavior hence the hypothesis
   that there must be huge amounts of nonluminous matter hidden in the halo. This
   unseen matter is given the technical name 'dark matter', but so far
   no conclusive
 experimental evidence is obtained for the existence of dark matter .

            Milgrom\cite{mm} proposed a new way of explaining galaxy
rotation curves by modifying the usual Newtonian dynamics. Such a
programme is called MOdified Newtonian Dynamics (MOND). In this
programme he introduced a new constant acceleration usually denoted
by '$a_{0}$', which has a value of about $10^{-10}m/s^{2}$. The MOND
approach can explain the rotation curves of galaxies as well as
those of clusters to a good approximation with out invoking the
existence of  dark matter. This theory can explain the gravitational
lensing also.

        In field theories, Hamiltonian can be derived from a covariant action.
In general relativity the situation is a little complicated by the
fact that the Einstein-Hilbert action includes surface terms and
generally
 the surface terms are ignored in the calculation. A new development in this
direction is to include surface terms in the calculations. Recently
Hawking derived Hamiltonian from Einstein-Hilbert  action retaining
all surface terms\cite{hw}. With the help of such surface terms
Easson et al. got a modified Friedmann equation and they were able
to explain the recently observed acceleration of the
universe\cite{ea}.

    In this paper we study the classical limit of modified Friedmann
equation derived by Easson et.al\cite{ea}.We get a modified force
equation and this equation is applied to a galactic scale. In the
present calculation two terms are appearing in the expression for
the gravitational force, one depending on the inverse of R and other
depending on inverse of $R^2$. The inverse of $R^2$ term is always
attractive but the first term ( which depends on inverse of R) can
be  attractive or repulsive. We study both cases and their
consequences.

We start with the covariant Lorentzian action with surface terms as
introduced by Hawking\cite{hw},
\begin{equation}
I(g,\Phi )=\int_{M}[\frac{\tilde{R}}{16\pi }+\mathcal{L}(g,\Phi
)]+\frac{1}{8\pi }\oint K \label{1},
\end{equation}

where $\tilde{R}$ is the Ricci scalar, $\mathcal{L}$ is the matter
Lagrangian density  and K is the trace of the extrinsic curvature of
the boundary, from which Friedman equation can be derived \cite{ea}:
\begin{equation}
\frac{\ddot{a}}{a}=-\frac{4\pi G}{3}(\rho +3P)+\frac{3}{2\pi }H^{2}+\frac{3}{%
4\pi }\dot{H}  \label{1},
\end{equation}

where 'a' is the cosmological scale factor, H the Hubble constant
$\rho$ and P are cosmic density and pressure respectively. Assuming
pressure to be zero for this case the above equation becomes

\begin{equation}
c_{1}\frac{\ddot{a}}{a}=-\frac{4\pi G}{3}\rho
+c_{2}(\frac{\dot{a}}{a})^{2},
\end{equation}
where $c_{1}$ and $c_{2}$ are two constants  Multiplying with 'a' on
both sides and again with 'r' and assuming $ar=R$\cite{as}, we find
\begin{equation}
\ddot{R}=-\alpha \frac{4\pi G\rho R}{3}+\beta \frac{\dot{R}^{2}}{R},
\end{equation}
where $\alpha$ and $\beta$ are two constants. From the above
equation we can write the equation for acceleration as,
\begin{equation}
\ddot{R}=-\alpha \frac{GM}{R^{2}}+\beta \frac{\dot{R}^{2}}{R}.
\end{equation}
Assuming $F=m\ddot{R}$ and $\dot{R}=V_{e}$ the expansion velocity of
universe, which is added to the equation of force otherwise similar
to Newtons law. The final equation of force can be written as
\begin{equation}
F=-\alpha \frac{GMm}{R^{2}}+\beta m\frac{V_{e}^{2}}{R}.
\end{equation}
This is similar to Newton's law but, a new term  appears which is
similar to a centripetal acceleration which depends on cosmic
expansion rate which will be very small. A force proportional to
$R^{-1}$  occurs in (MOND)\cite{pp}. This shows that expansion of
the universe affects the normal motion with a very small force whose
acceleration looks like a centripetal acceleration. But if it is
repulsive there may be a possibility that the two terms in the right
hand side may cancel each other  and can have a force free region
and the value of R for this case is given by
\begin{equation}
R=\frac{\alpha GM}{\beta V_{e}^{2}}.
\end{equation}
    This is contradictory to our common experience. But if it is
attractive it gives a correction to Newton's law. In the galactic
scale, the second term in equation (6) dominates and give constant
rotational velocity
\begin{equation}
V=V_{e},
\end{equation}
thus explaining the flat rotation curves of $V_{e}$.

We now take the correction term as an attractive one. The correction
term gives a constant acceleration. If we calculate using R as the
radius of cosmic horizon, $R=c/H$ then the correction term becomes
$\beta cH$. From the studies of MOND we know the asymptotic
acceleration is cH. If $\beta =1$ this gives the MOND acceleration
term, then the equation can be written as
\begin{equation}
F=-\frac{\alpha GMm}{R^{2}}+\beta mcH.
\end{equation}
The second term gives the asymptotic acceleration cH of MOND
\cite{pp}.

The gravitational potential also gets modified in this case. The
modified potential is given by (for a test mass m )
\begin{equation}
F=-\frac{\partial \Phi }{\partial R},
\end{equation}
and is given by
\begin{equation}
\Phi =-\frac{\alpha GMm}{R}+\frac{\beta mV_{e}^{2}} {2}.
\end{equation}
The potential depends on the cosmic expansion rate from which we can
also calculate the modified escape velocity  which is given by,
\begin{equation}
v=\sqrt[2]{\frac{2\alpha GM}{R}-\beta V_{e}^{2}}.
\end{equation}
This equation shows that the escape velocity depends on cosmic
expansion rate.

The Newtonian equation for planetary motion is
\begin{equation}
F=m[\frac{d^{2}R}{dt^{2}}-R(\frac{d\theta }{dt})^{2}],
\end{equation}
and in the present case this equation gets modified as,
\begin{equation}
m[\frac{d^{2}R}{dt^{2}}-R(\frac{d\theta }{dt})^{2}]=-\frac{\alpha GMm}{R^{2}}%
+\frac{\beta mV_{e}^{2}}{R}.
\end{equation}
This equation can be simplified as
\begin{equation}
\frac{d^{2}R}{dt^{2}}-R(\omega ^{2}+\beta \omega _{e}^{2})=-\frac{\alpha GM}{%
R^{2}},
\end{equation}
where $\omega_{e}$ is the angular velocity derived from cosmic
expansion. If we call
\begin{equation}
(\omega ^{2}+\beta \omega _{e}^{2})=\omega _{tot}^{2}.
\end{equation}
 the equation of planetary motion now takes the normal form
\begin{equation}
\frac{d^{2}R}{dt^{2}}-R\omega _{tot}^{2}=-\alpha \frac{GM}{R^{2}}.
\end{equation}
This leads to the  usual elliptical motion and is similar to
Newtonian case. But here the correction appears in the angular
velocity term and when the angular velocity $\omega$ is less than
$\omega_{e}$ the second term dominates. Since it is very small its
effect appears only to slowly moving distant planets.

Introduction of surface term modifies Friedman equation and in the
classical limit we get a Newtonian analogue equation. If the
correction term is attractive then it gives a result as in the case
of MOND. If it is a repulsive term there exists a force free
distance from the source in which gravitational attractive force
exactly cancels the repulsive force. The modified equation leads to
the law of planetary motion for normal angular velocities of the
planet and gets modified considerably for very small angular
velocities. The effect of second term in equation (6) dominates in
galactic scale. It gives constant rotational velocities for stars in
the halo region and thus explains the flat rotation curves of stars
and galaxies

SS wishes to thank CSIR for providing fellowship under CSIR JRF
scheme. VCK acknowledges Associateship of IUCAA, Pune

\end{document}